\documentclass{aa}

\usepackage{amssymb}
\usepackage{graphicx}
\usepackage{pstricks}
\usepackage{txfonts}

\usepackage{natbib}

\begin{document}

\title{Spin-Orbit angle distribution \\
and the origin of (mis)aligned hot Jupiters}

\titlerunning{Spin-Orbit angle distribution}

\author{A. Crida \inst{1}
  \and K. Batygin \inst{2}}
\institute{Laboratoire Lagrange, UMR7293, Universit\'e Nice Sophia-antipolis / CNRS / Observatoire de la C\^ote d'Azur, 06300 Nice, \textsc{France}
\\ \email{crida@oca.eu}
\and Institute for Theory and Computation, 
Harvard-Smithsonian Center for Astrophysics,
60 Garden St., Cambridge, MA 02138
}

\date{Received December 19, 2013 / Accepted April 7, 2014}

\abstract{
For 61 transiting hot Jupiters, the projection of the angle between the orbital plane and the stellar equator (called the spin-orbit angle) has been measured. For about half of them, a significant misalignment is detected, and retrograde planets have been observed. This challenges scenarios of the formation of hot Jupiters.
}{
In order to better constrain formation models, we relate the distribution of the real spin-orbit angle $\Psi$ to the projected one $\beta$. Then, a comparison with the observations is relevant.
}{
We analyse the geometry of the problem to link analytically the projected angle $\beta$ to the real spin-orbit angle $\Psi$. The distribution of $\Psi$ expected in various models is taken from the literature, or derived with a simplified model and Monte-Carlo simulations in the case of the disk-torquing mechanism.
}{
An easy formula to compute the probability density function (PDF) of $\beta$ knowing the PDF of $\Psi$ is provided. All models tested here look compatible with the observed distribution beyond 40 degrees, which is so far poorly constrained by only 18 observations. But only the disk-torquing mechanism can account for the excess of aligned hot Jupiters, provided that the torquing is not always efficient. This is the case if the exciting binaries have semi-major axes as large as $\sim 10^4$~AU.
}{
Based on comparison with the set of observations available today, scattering models and the Kozai cycle with tidal friction models can not be solely responsible for the production of all hot Jupiters. Conversely, the presently observed distribution of the spin-orbit angles is compatible with most hot Jupiters having been transported by smooth migration inside a proto-planetary disk, itself possibly torqued by a companion.
}


\maketitle

\section{Introduction}

The existence of close-in giant planets whose orbits lie in close
proximity to their host stars is now well established
\citep[e.g.,][]{Cumming-2011}. Although these objects constitute the
best observationally characterised sample of exoplanets, their origins
remain puzzling from a theoretical point of view. As in-situ formation
of hot Jupiters is problematic \citep{Chiang-Laughlin-2013}, it is
likely that these objects have formed beyond the ice-line in the
proto-planetary disk (i.e, at an orbital separation of a few AU) and
have since been transported inwards
\citep{Lin-etal-1996}. Traditionally, smooth migration forced by
interaction with the proto-planetary disc in which planets form, has
been invoked to facilitate transport
\citep{LinPapaloizou1986b,Lin-etal-1996,Wu-Murray-2003,Crida-Morby-2007,Papaloizou-etal-2007}. Recently,
substantially more violent processes have been proposed to account for
the generation of hot Jupiters
\citep{Fabrycky-Tremaine-2007,Nagasawa-etal-2008,Beauge-Nesvorny-2012}.
However, the dominance of the roles played by each mechanism remains
controversial \citep{Dawson-etal-2012arXiv}. Accordingly, observations
of the stellar spin-planetary orbit misalignment has been invoked as a
means of differentiating among the proposed models.

Among the $\sim 1000$ exoplanets detected to date, $\sim 60$ have an
observed measure of the angle between their orbital plane and the
equatorial plane of their host star
\citep[e.g.,][]{Winn-etal-2007,Triaud-etal-2010}. Traditionally, this
has been achieved through the Rossiter-McLaughlin effect
\citep{McLaughlin-1924}\,; however, recently novel techniques involving
asteroseismology \citep{Huber-etal-2013} and star-spot measurements
\citep{Hirano-etal-2012} have also been utilised to this end. A wide
variety of measured angles have been reported to date, and many
exo-planets are apparently characterised by \emph{spin-orbit
  misalignment}\,: the orbital obliquity differs significantly from
$0$. Instinctively, this is surprising as planets are supposed to form
within a proto-planetary disk whose angular momentum direction is the
same as that of the star. Therefore, the origin of the spin-orbit
misalignment has received considerable attention over the past few
years.

A promising way to disentangle the roles of the various transport
mechanisms, is to analyse the observed distribution of the spin-orbit
angles, and to compare it to the expected distribution from a given
model. However, observations only provide the \emph{projected}
spin-orbit angle, not the real one. Accordingly, in
section~\ref{sec:proj}, we describe the 3D geometry of the problem,
and we infer the distribution of the projected spin-orbit angles from the
real ones (and vice-versa).

In section~\ref{sec:PDFs}, we compute the distributions of the
projected spin-orbit angles expected from various mechanisms, devoting
special attention to the disk-torquing model
\citep{Batygin-2012}. Within the framework of the disk-torquing
mechanism, a (possibly transient) companion to a young star makes the
proto-planetary disk precess around the binary's axis, so that the
plane in which planets eventually form and the equatorial plane of the
central star can differ. Contrary to the violent category of migration
mechanisms --\,e.g., planet-planet scattering
\citep{Rasio-Ford-1996,Ford-Rasio-2008} and Kozai resonance
\citep{Wu-Murray-2003,Fabrycky-Tremaine-2007,Naoz-etal-2011}\,-- this
process predicts that all the planets of the system can share the same
misalignment. The recent discovery of significant misalignment among a
multi-transiting system \citep{Huber-etal-2013} therefore puts
emphasis on this model.

Using analytical arguments and Monte-Carlo simulations, we find that
the current observational aggregate can be well explained by the
disk-torquing effect, implying that disk-driven planet migration in
(torqued) disks could be the dominant source of (mis)aligned hot
Jupiters.

\section{The 3D geometry of the spin-orbit angle}
\label{sec:proj}

There is considerable confusion in the literature about the spin-orbit
angle. Let us define precisely what we are interested in here. The
true misalignment angle is actually the angle between two vectors in
3D space\,: $\vec{L}_p$, the orbital angular momentum of the planet,
and $\vec{L}_s$, the angular momentum of the spin of the star. As
such, it can only lie between $0$ and $180$ degrees (there are no
negative angles in 3D). This real, 3D angle is denoted below as
$\Psi$.

The projections of these two vectors onto the plane of the sky (noted
$\vec{L}_p'$ and $\vec{L}_s'$) form an oriented angle, which in
principle could range between $-180^\circ$ and $+180^\circ$. However,
whether the angle ($\vec{L}_p',\vec{L}_s'$) is positive or negative
when measured clockwise, corresponds to the same 3D configuration,
observed from either side of the star, from the ascending or
descending node. In any case, the two cases are indistinguishable
observationally (Triaud, personal communication). Thus, it does not
make sense to report negative angles. The projected spin-orbit angle
should also be expressed as between $0^\circ$ and $180^\circ$. Still,
on \texttt{exoplanets.org}, one finds many negative projected
spin-orbit angles, taken from published articles. This angle is
sometimes noted $\lambda$
\citep[e.g.,][\texttt{exoplanets.org}]{Fabrycky-Winn-2009}, and
sometimes it is noted as $\beta$ \citep[e.g.,][]{Triaud-etal-2010}. The
projected angle is denoted below as $\beta$, and will be taken as the
absolute value of the misalignment angle $\beta$ or $\lambda$ reported
in the literature.

\subsection{Relation between the real and projected spin-orbit angle}

Assume that $\Psi$ is fixed. Which $\beta$ will be observed~? What is
the probability density function (PDF) of $\beta$~?

This question has already been addressed in general by
\citet{Fabrycky-Winn-2009}. They provide $\beta$ (that they note
$\lambda$) as a function of $\Psi$ and $i_o$, the inclination of the
orbital angular momentum with respect to the line of sight. Here, we
propose a simpler, one column derivation, making the assumption that
the observer is exactly in the orbital plane ($i_o=\pi/2$)\,; this is
appropriate because all the planets with known spin-orbit angle
transit their host star, so $|\pi/2-i_o|$ is very small. In this case,
$\vec{L}_p=\vec{L}_p'$ is perpendicular to the line of sight, and is
in the plane of the sky. Let us consider spherical coordinates
($r,\phi,\theta$) centred on the star, such that colatitude $\theta=
0$ corresponds to the north pole of the orbit, the direction of
$\vec{L}_p$\,; the origin of the longitude ($\theta=90^\circ$,
$\phi=0$) corresponds to the direction of the observer. In these
coordinates, $\vec{L}_s$ has a colatitude $\theta_s=\Psi$. Its azimuth
$\phi_s$ \footnote{This angle $\phi_s$ was noted $\Omega$ by
  \citet{Fabrycky-Winn-2009}} can be anything between $0$ and $2\pi$,
with a uniform distribution. Clearly, if $\phi_s\equiv 0 [\pi]$, the
observer sees $\beta=0$ if $\Psi<\pi/2$, and $\beta=\pi$ if
$\Psi>\pi/2$\,; conversely, if and only if $\phi_s=\pm\pi/2$ the
observer sees $\beta=\Psi$ exactly.

\begin{figure}
\includegraphics[width=\linewidth]{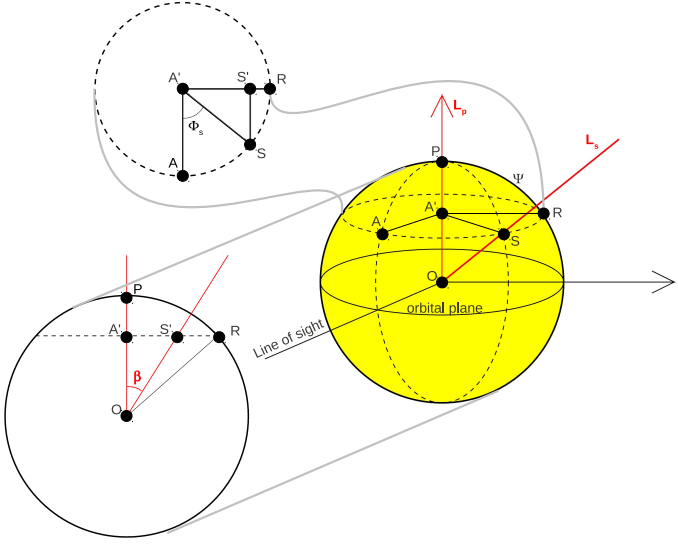}
\caption{\textbf{Right\,:} 3D representation of the problem. The yellow sphere is the unit sphere centred on the star. P marks the direction of the orbital angular momentum vector $\vec{L}_p$ and S that of the stellar spin $\vec{L}_s$. The dashed circle passing through points A and S gathers all the points making an angle $\Psi$ with P. It is represented in the top left.\newline
\textbf{Top left\,:} The circle of the unit sphere gathering all the points at colatitude $\Psi$ with respect to the orbital angular momentum vector of the planet. A is the point facing the observer\,; S is the point corresponding to the direction of the stellar spin. A and S are projected on the diameter of this circle perpendicular to the line of sight onto A' and S'\,; $\phi_s$ is then $\widehat{AA'S}$.\newline
\textbf{Bottom left\,:} The projected plane, as seen by the observer. The previous dashed circle is now a dashed horizontal line, on which A' and S' are the projections of A and S along the direction of the line of sight. The arc PR defines an angle $\Psi$, while the projected spin orbit angle $\beta$ is $\widehat{A'OS'}$, marked in red.
}
\label{fig:proj}
\end{figure}

This configuration is shown in the right panel of
Figure~\ref{fig:proj}. The point $O$ is the centre of the unit sphere, and $P$ and $S$
are the intersection of the sphere with the vectors $\vec{L}_p$ and
$\vec{L}_s$ respectively. The point $A$ has colatitude $\theta=\Psi$
and azimuth $\phi=0$, facing the observer, while $R$ has colatitude
$\theta=\Psi$ and azimuth $\phi=\pi/2$, appearing on the limb of the
star for the observer. Accordingly, $OA=OS=OP=OR=1$\,.

The plane $(OPR)$ is the plane perpendicular to the line of sight
passing through $O$\,: it is the plane of the sky as seen by the
observer, onto which everything is orthogonally projected, along the
direction of the line of sight.

The circle gathering all the points with colatitude $\Psi$ is the
dashed circle passing through $A$, $S$, and $R$. This circle is represented
in the top left of Figure~\ref{fig:proj}. The orthogonal projections
of $A$ and $S$ onto the plane $(OPR)$, are $A'$ and $S'$ respectively. The
centre of this dashed circle is $A'$, and its radius is obviously
$A'A=A'S=A'R=\sin\Psi$. Thus, $A'S'=\sin\Psi\,|\sin\phi_s|$.

In the projected plane (shown in bottom left of Fig.~\ref{fig:proj}),
the angle between the north pole of the orbit and the spin of the star
appears to be $\beta=\widehat{POS'}$. As $A'$ is the orthogonal
projection of $S'$ on the $(OP)$ line, we have $\tan\beta=A'S'/OA'$, where
$OA'=\cos\Psi$ is negative when $\Psi>\pi/2$. Finally,
\begin{eqnarray}
\label{eq:beta1}
\beta & = & \arctan\left(|\sin\phi_s|\,\tan\Psi\right) \equiv G(\phi_s)\\
|\phi_s| & = & \arcsin\left(\frac{\tan\beta}{\tan\Psi}\right)\ .
\end{eqnarray}
This is equivalent to Eq.~(11) of \citet{Fabrycky-Winn-2009}, with $i_o=\pi/2$, $\Omega=\phi_s$ and $\lambda=\beta$.

\subsection{Probability density function of $\beta$, for fixed $\Psi$}

Now, as the distribution of $\phi_s$ is uniform in the interval $[0;2\pi[$,
and $|\sin(x)|=|\sin(\pi-x)|=|\sin(\pi+x)|=|\sin(2\pi-x)|$, it is
sufficient to consider a uniform distribution for $0\leqslant\phi_s<\pi/2$, with
probability density $2/\pi$. In this case, $\beta$ is a monotonic
function of $\phi_s$. It is well known that if $X$ is a random variable
of probability density function $f_X$, and $Y=G(X)$ with G a monotonic
function, then the PDF of $Y$ is
\begin{equation}
f_Y(y) = f_X\left(G^{-1}(y)\right)\times \left|\left(G^{-1}\right)'(y)\right|\ .
\label{eq:compoVA}
\end{equation}

Thus, using Eq.~(\ref{eq:beta1}) for fixed $\Psi$, the PDF of $\beta$ is\,:
\begin{equation}
f(\beta|\Psi)=\left\{\begin{array}{ll}
\displaystyle\frac{2}{\pi}\frac{1+\tan^2\beta}{(\tan^2\Psi-\tan^2\beta)^{1/2}}& {\rm if\ }\beta\in\mathcal{T}\ ,\vspace{3pt}\\
0 & {\rm otherwise}
\end{array}\right.
\label{eq:project1}
\end{equation}
where $\mathcal{T}=\{ 0\leqslant\beta<\Psi<\frac{\pi}{2}\}
\cup \{\frac{\pi}{2}<\Psi<\beta\leqslant\pi\}$.

This equation is identical to Eq.~(19) of \citet{Fabrycky-Winn-2009}\footnote{Using $1+\tan^2(u)=1/\cos^2(u)$ leads to their expression easily.}.

\begin{figure}
\includegraphics[width=\linewidth]{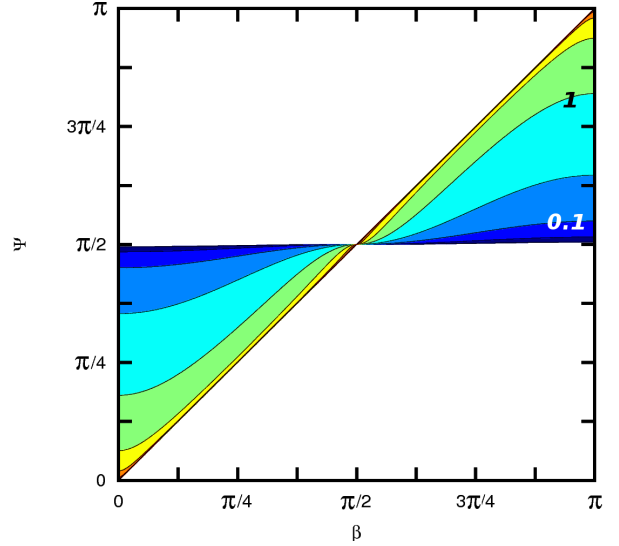}
\caption{Contour map of $f(\beta|\Psi)$ as defined by
  Eq.~(\ref{eq:project1}). The contours correspond to values $10^k$,
  with $k=-1.5,\,-1\ldots 1,\,1.5$\,; the curves corresponding to
  $f=1$ and $f=0.1$ are marked. Note that $f$ actually diverges
  towards $+\infty$ approaching the $\Psi=\beta$ line.  }
\label{fig:f_Psi_beta}
\end{figure}

Figure~\ref{fig:f_Psi_beta} shows the decimal logarithm of
$f(\beta|\Psi)$ in the $\beta-\Psi$ plane. For a given real spin-orbit
angle $\Psi_0$, the PDF of the observed projected angle $\beta$ can be
found by going along a $\Psi=\Psi_0$ line in the figure. One can check
analytically that for all $\Psi$,
$\displaystyle\int_{\beta=0}^{\beta=\pi}f(\beta|\Psi)\,{\rm
  d}\beta=1$~, as it should.

\subsection{Conversion of the PDF of $\Psi$ into the PDF of $\beta$}

If now $\Psi$ has its own PDF $w(\Psi)$, the corresponding PDF of
$\beta$ will be\,:
\begin{equation}
f(\beta)=\int_{\Psi=0}^{\Psi=\pi}f(\beta|\Psi)w(\Psi)\ {\rm d}\Psi\ .
\label{eq:f_b}
\end{equation}
Computing this integration corresponds to summing vertically in
Figure~\ref{fig:f_Psi_beta}, after having given to every horizontal
line a weight $w(\Psi)$.

\begin{figure}
\includegraphics[width=\linewidth]{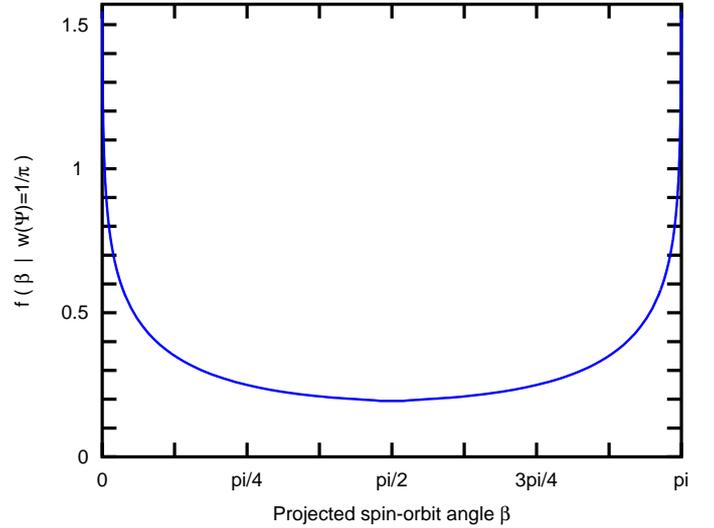}
\caption{Representation $f(\beta)$ when $w(\Psi)=1/\pi$ (uniform
  distribution of $\Psi$).}
\label{fig:horny}
\end{figure}

For example, assuming a uniform $w(\Psi)=1/\pi$, one gets the
double-peaked distribution shown in Fig.~\ref{fig:horny}. On the other
hand, assuming that $\Psi$ is isotropically distributed,
$w(\Psi)=(\sin{\Psi})/2$, one can solve Eq.~(\ref{eq:f_b}) and find
$f(\beta)=1/\pi$, uniform, as expected. The observed distribution of
$\beta$ (see histogram in Fig.~\ref{fig:Bat-anal}) is neither flat nor
reminiscent of the distribution shown in Fig.~\ref{fig:horny},
implying that the real distribution of $\Psi$ is neither uniform, nor
isotropic.

\subsection{Deprojection}

Let $i_s$ be the angle between the stellar angular momentum $\vec{L_s}$
and the line of sight\,; an isotropic distribution of $\vec{L_s}$ on
the unit sphere gives to $i_s$ a PDF $f(i_s)=\frac12\sin(i_s)$ between
$0$ and $\pi$. On the projected plane (bottom left of figure~1), we
now have $OS'=\sin i_s$, and
$OA'=OS'\times\cos\beta=OR\times\cos\Psi=\cos\Psi$. Thus,
\begin{equation}
\Psi=\arccos(\cos\beta\times\sin i_s) = G_2(i_s)\ .
\label{eq:Psi}
\end{equation}
Note that $i_s$ or $\pi-i_s$ give exactly the same projection. One can
therefore assume that $i_s$ is distributed between $0$ and $\pi/2$
with PDF $f(i_s)=\sin(i_s)$. Then, using Eqs.~(3) and
(\ref{eq:Psi}), one gets\,:
\begin{equation}
f(\Psi|\beta) = \left\{\begin{array}{ll}
\vspace{6pt}\frac{\cos\Psi}{\cos\beta}\frac{\sin\Psi}{\sqrt{\cos^2\beta-\cos^2\Psi}}
& \begin{array}{l}{\rm if\ }\ 0<\beta<\Psi<\pi/2\\{\rm or\ }\pi/2<\Psi<\beta<\pi\ ,\end{array}\\
0 & \begin{array}{l}{\rm otherwise\ ,}\end{array}
\end{array}\right.
\label{eq:DPF_Psi}
\end{equation}
in agreement with Eq.~(21) of \citet{Fabrycky-Winn-2009}.

Then, the PDF of $\Psi$ can be deduced from the observations of $\beta$\,:
\begin{equation}
f(\Psi)=\sum_{i} f(\Psi|\beta_i)
\label{eq:deproj}
\end{equation}
where the index $i$ spans the whole sample.

\begin{figure}
\includegraphics[width=\linewidth]{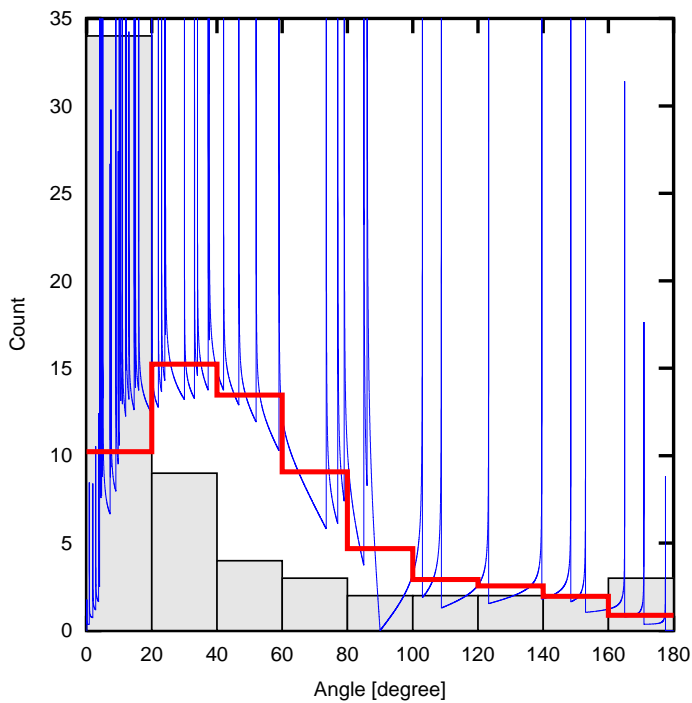}
\caption{Grey shaded histogram\,: distribution of the observed projected spin-orbit angle $\beta$ (taken as $|\beta|$ or $|\lambda|$ in the data from \texttt{exoplanets.org}). Blue thin curve\,: $f(\Psi)$, as given by Eq.~(\ref{eq:deproj}). Red stairs\,: Histogram of the PDF of $\Psi$ with bins of $20^\circ$.}
\label{fig:deproj}
\end{figure}

Applying this to the data found on \texttt{exoplanets.org} at the end of 2013,
we get an irregular curve peaked at each of the $\beta_i$, as $f(\Psi|\beta_i)$ diverges towards $+\infty$ as $\Psi\to\beta_i$. This curve is displayed as a thin blue line in Figure~\ref{fig:deproj}. To avoid this, one could smooth the data using the error-bars $\sigma_i$ and compute\,:
$$f(\Psi)=\sum_{i} \int \frac{1}{\sqrt{2\pi}\sigma_i}\exp\left(-\frac{(\beta-\beta_i)^2}{2\,{\sigma_i}^2}\right)\ f(\Psi|\beta)\ {\rm d}\beta\ .$$
However, the $\sigma_i$ are so diverse (from $0.3$ to $60^\circ$) that this operation would only degrade the information, without completely smoothing the curve. Therefore, we prefer to operate similar to the observations of $\beta$ and build a histogram\,: we average $f(\Psi)$ on successive intervals of $20^\circ$ width. This is represented in Figure~\ref{fig:deproj} as the red stairs, while the histogram of the observations of $\beta$ is shaded grey.
It seems here that aligned hot Jupiters are a minority. However, one should realize that if all hot Jupiters were actually aligned, $\Psi=\beta=0$, but Eq~(\ref{eq:DPF_Psi}) gives $f(\Psi|\beta=0)=\textrm{max}\{\cos\Psi\,,\,0\}$\,; it is therefore almost impossible that the first step of the red stairs is the highest step, even if there is a large majority of aligned hot Jupiters.

In any case, with only $18$ cases with $\beta>40^\circ$ in the data, and
large error bars, the statistics is rather poor to infer the PDF of
$\Psi$ accurately. Therefore, in what follows, we prefer to infer the
PDF of $\beta$ from various mechanisms, and compare it directly with
the observed distribution of $\beta$.

\section{Application to proposed mechanisms}
\label{sec:PDFs}

In this section, we compute/take the PDF of $\Psi$ that is expected
from several mechanisms of formation of hot Jupiters. Then, we compute
the corresponding PDF of $\beta$ using Eq.~(\ref{eq:f_b}), and
compare it with the distribution of the observations.
The distribution of the observations is shown as a grey shaded histogram in Figures~\ref{fig:Bat-anal}, \ref{fig:Bat-MCall}, and \ref{fig:All}. The bins have a width of $20^\circ$, and many of them contain only two cases. This small number statistics is prone to significant variations with new observations, or change of the bins\,: one object more or less in a bin represents a $50\%$ variation. Nonetheless, it appears robust that the distribution of the projected spin-orbit angle beyond $60^\circ$ is almost uniform. Models should account for this.

To be more precise, assuming the small number of planets in our bins follows a Poisson process of parameter $\lambda$, when $N$ are found in a bin, the likelihood of $\lambda$ is given by $\lambda^N\exp(-\lambda)/N!$~. Thus, in the bins where two are found, one can say with $50\%$ confidence that $1.17<\lambda<3.16$ and with $95\%$ confidence that $0.31<\lambda<6.34$.

\subsection{Disk torquing}

Contemporary observational surveys suggest that a considerable
fraction of solar-type stars are born as binary or multiple systems
\citep{Ghez-etal-1993,Kraus-etal-2011}. Moreover, most stars form in
embedded cluster environments \citep{Lada-Lada-2003} where dynamical
evolution can lead to the acquisition of transient companions
\citep{Malmberg-etal-2007}. Recently, \citet{Batygin-2012} showed that
the presence of a companion to a young star can force the
proto-planetary disk to precess around the binary's axis, so that the
plane in which planets eventually form and the equatorial plane of the
central star can differ.

\subsubsection{Analytic simple model}

Denoting the angle between the orbital plane of the companion and
the stellar equator as $i\,'$ (therefore, $0\leqslant i\,' \leqslant \pi/2$),
$\Psi$ evolves between $0$ and $2i\,'$ in this mechanism. At first sight
of Fig.~2 of \citet{Batygin-2012}, it seems reasonable to consider
this evolution as linear with time, back and forth. In the end, the
PDF of $\Psi$ is a uniform distribution between $0$ and $2i\,'$\,:
\begin{equation}
f(\Psi\,|\,i\,')=\frac{1}{2i\,'}\ \textrm{\ if\ }0\leqslant\Psi\leqslant2i\,'\ ;\ \ 0 \textrm{\ otherwise.}
\label{eq:fPi'}
\end{equation}
The PDF of $i\,'$ should correspond to an isotropic distribution of the
orbital angular momentum vector of the companion with respect to the
stellar spin, thus it reads\,:
\begin{equation}
g(i\,') =  \sin(i\,')\ .
\label{eq:g}
\end{equation}

\begin{figure}
\includegraphics[width=\linewidth]{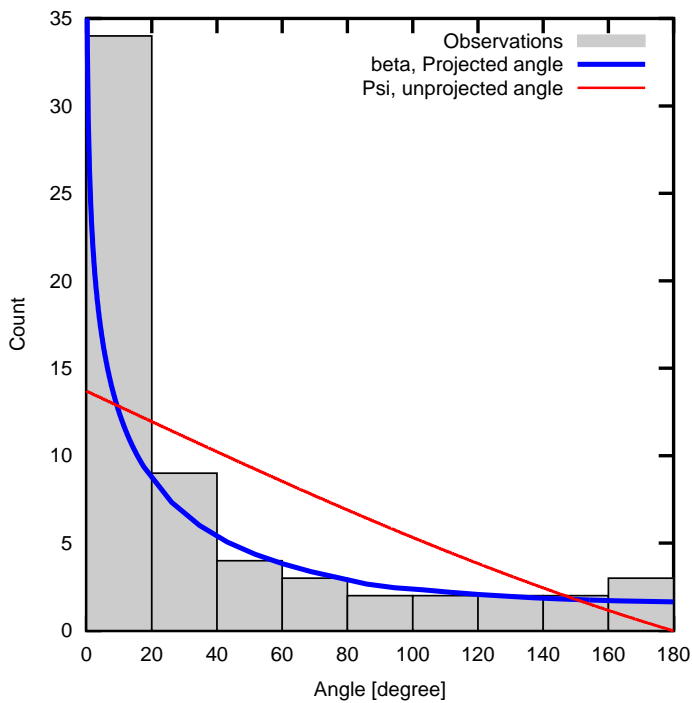}
\caption{Distribution of the spin - orbit angle expected in the
  disk-torquing mechanism. Red thin line\,: distribution of $\Psi$,
  given by Eq.~(\ref{eq:f-de-Psi}). Blue thick line\,: corresponding
  PDF of $\beta$. Background histogram\,: observations.}
\label{fig:Bat-anal}
\end{figure}

Now, the PDF of $\Psi$ is given by\,:
\begin{eqnarray}
f_{B12}(\Psi) & = & \int_0^{\pi/2}f(\Psi\,|\,i\,')g(i\,')\ {\rm d}i\,'\nonumber\\
 & = & \int_{\Psi/2}^{\pi/2} \frac{\sin(i\,')}{2i\,'}\,{\rm d}i\,'\nonumber\\
f_{B12}(\Psi)  & = & \frac{1}{2}\left[\textrm{Si}(\pi/2)-\textrm{Si}(\Psi/2)\right]
\label{eq:f-de-Psi}
\end{eqnarray}
where $\mathrm{Si}(x)=\int_0^x\frac{\sin(t)}{t}\,{\rm d}t$, which has
unfortunately no easy analytical expression. Nonetheless, this
expression can be computed numerically. On Figure~\ref{fig:Bat-anal},
$f_{B12}$ is displayed as the red thin curve\,; it looks linear, but
it's actually not a straight line. The corresponding PDF of $\beta$
(using Eq.~(\ref{eq:f_b})\,) is shown as the blue thick curve. The
major difference between the two curves enlights the necessity of
taking the projection into account. In the background, the histogram
of the observations shows a good agreement with the predicted
distribution of the projected angles. The PDF of $\beta$ is normalised
to have $18$ planets with $\beta>40^\circ$, like the data.

With this normalisation, the number of planets with $\beta<20^\circ$
(resp. $20^\circ<\beta<40^\circ$) is only $14.2$ (resp. $6.8$), while
$34$ (resp. $9$) are observed. This excess of aligned planets in the
observations should not bother us. Indeed, hot Jupiters form in the
proto-planetary disk, and migrate inwards irrespective of the
multiplicity of the stellar system. Thus, if the proto-planetary disk
is not torqued (e.g., the star is always single), the hot Jupiters are
expected to be aligned. If the proto-planetary disk is torqued by a
companion, then the planets will be misaligned, with the distribution
given above. The apparent excess of aligned hot Jupiter can therefore
be interpreted as the fraction of disks that were never significantly
torqued. We come back on this issue in the next sub-section.

\subsubsection{Monte Carlo simulations}

Within the framework of the model proposed by \citet{Batygin-2012}, the stellar spin axis is taken to remain in the primordial plane of the disk for all time. Physically, this simplifying assumption corresponds to a non-accreting, unmagnetised young stellar object that rotates at a negligibly slow rate. This picture is somewhat contrary to real pre-main-sequence stars, which typically accrete $\sim 10^{-8} M_{\odot}$/year from the disk \citep{Hartmann-2008,Hillenbrand-2008}, have $\sim 1$ kGauss magnetic fields at the stellar surface \citep{Shu-etal-1994a,Gregory-2010}, and rotate with characteristic periods in the range $P_{\rm{rot}}\simeq 1-10$~days \citep{Herbst-etal-2007}. Accordingly, \citet{Batygin-Adams-2013arXiv} examined magnetically and gravitationally facilitated disk-star angular momentum transfer with an eye towards constraining the conditions needed for the acquisition of spin-orbit misalignment. They showed that the excitation of spin-orbit misalignment is only quenched when the host star continuously spins up because of gravitational contraction \citep[i.e., the stellar field is too weak for magnetic breaking to occur\,; see][]{Shu-etal-1994a,Matt-Pudritz-2005ApJ,Matt-Pudritz-2005MNRAS}. However, they also found that the process by which spin-orbit misalignment is attained is somewhat more complicated than that described in \citet{Batygin-2012}. Specifically, as the disk mass decreases throughout its lifetime, the gravitational coupling between the star's quadrupole moment (that arises from rotational deformation) and the torqued disk gives rise to orbital obliquity via a secular resonance encounter between the stellar spin-axis precession frequency and the disk-torquing frequency \citep[see][for details]{Batygin-Adams-2013arXiv}.

\begin{figure}
\begin{center}
\includegraphics[width=\linewidth]{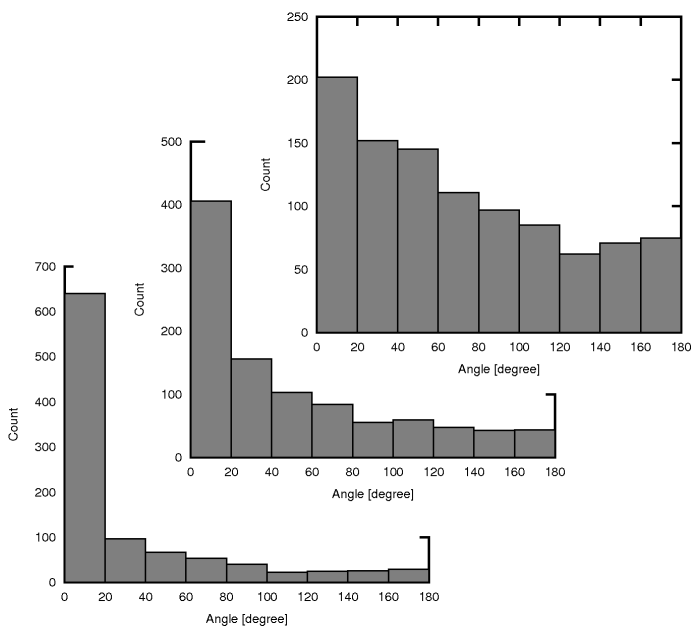}
\end{center}
\caption{Distributions of the projected spin - orbit angle found in Monte Carlo simulations as described in the text, for three values of the maximum of the log-flat distribution of the semi-major axis of the binary. Top right\,: $a_{\rm max}=10^3$~AU. Middle\,: $a_{\rm max}=10^{3.75}$~AU. Bottom left\,: $a_{\rm max}=10^{4.5}$~AU.}
\label{fig:Bat-MC}
\end{figure}

Cumulatively, the more complete model of \citet{Batygin-Adams-2013arXiv} does not easily lend itself to analytic manipulation. As a result, to derive the associated distribution of projected misalignment angles, we perform a Monte Carlo simulation, utilising their perturbative model (for brevity, we shall not rehash their formalism here, but instead refer the reader to their description). As above, the inclinations of the binary companion stars are taken to be isotropic, the binary semi-major axis is drawn from a log-flat distribution spanning $10^{2.5}-a_{\rm max}$~AU \citep[where $a_{\rm max}$ can vary, see caption of Figure~\ref{fig:Bat-MC}\,;][]{Kraus-etal-2011}, while the eccentricity and the primary-to-perturber mass ratio are taken to be uniform in the intervals [0,1] and [0.1,10], respectively \citep[see][]{Kraus-etal-2008}. The primary star's mass and the disk's  initial mass are also drawn from uniform distribution spanning [0.5,1.5] $M_{\odot}$ and [0.01,0.05] $M_{\odot}$, respectively \citep{Herczeg-Hillenbrand-2008}. For all simulations, a surface density profile of the form $\Sigma \propto r^{-1}$ is assumed. The disk's outer edge is taken to lie between $10^{1.5}$ and $10^{2}$ AU \citep{Levison-etal-2008,Anderson-etal-2013}, while the inner edge corresponds to the stellar corotation radius \citep{Koenigl-1991,Shu-etal-1994a}. In turn, the stellar rotation periods are randomly chosen to lie between 1 and 10 days in rough agreement with the observational samples of \citet{Littlefair-etal-2010,Affer-etal-2013}. Following \citet{Gallet-Bouvier-2013}, pre-main sequence rotational evolution is ignored because of the inherent complexities. Finally, the disk mass loss, stellar structure, and stellar gravitational contraction are modelled as described in \citet{Batygin-Adams-2013arXiv}.

\begin{figure}
\includegraphics[width=\linewidth]{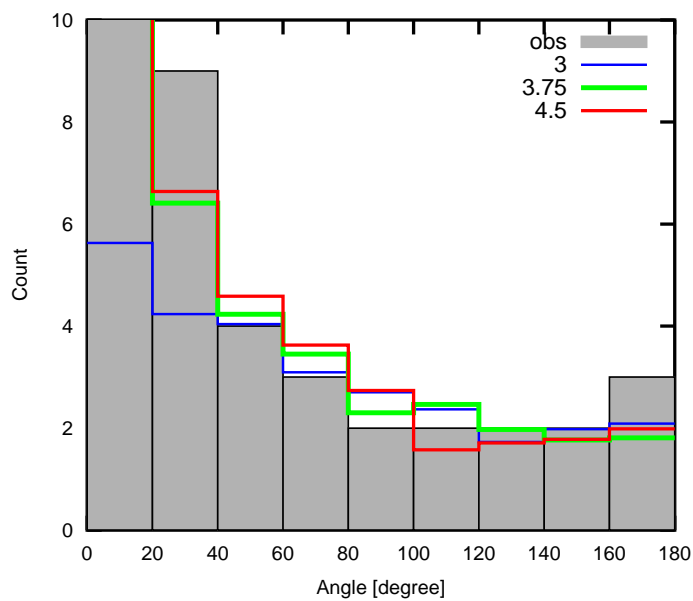}
\caption{Distributions produced by the Monte Carlo simulations, normalised to $18$ cases with $\beta>40^\circ$. Thin blue\,: $a_{\rm max}=10^3$~AU. Thick green\,: $a_{\rm max}=10^{3.75}$~AU. Red\,: $a_{\rm max}=10^{4.5}$~AU. Background histogram\,: observations.}
\label{fig:Bat-MCall}
\end{figure}

With the aforementioned ingredients in place, we compute the
disk-torquing frequency as described in \citet[][see equations 6 \& 7
  in the SI]{Batygin-2012}, while the rest of the calculation follows
directly from section 4 of \citet{Batygin-Adams-2013arXiv}. The
results are shown in Figure~\ref{fig:Bat-MC}. One thousand random sets
of parameters have been chosen with the distributions described above,
and the resulting $1000$ final spin - orbit angles have been binned in
bins of $20^\circ$ width to produce a histogram, to be compared to the
observations (Figure~\ref{fig:Bat-anal}). Three cases are presented,
where the only difference is the assumed value of $a_{\rm max}$, the
maximum possible value of the semi-major axis of the binary\,; from
left to right, $a_{\rm max}=10^3$, $10^{3.75}$, and $10^{4.5}$~AU. The
fraction of disks that will experience a negligible torquing appears
to be very sensitive to $a_{\rm max}$. However, the distribution of
the angles in the torqued cases remains unchanged, and close to the
observations, as can be seen in Figure~\ref{fig:Bat-MCall}. In this
figure, the three distributions obtained by the three sets of Monte Carlo
simulations have been normalised to have $18$ planets with
$\beta>40^\circ$, like in the data. They are represented in blue,
green, and red stairs, and are very similar beyond $40^\circ$. In the
background, the histogram of the data is shown in grey bars. All the
three stairs are within 1 count of the histogram, signaling excellent
agreement.

The choices of input parameters employed in the simulations are essentially naive estimates, predominantly guided by observational surveys. The observed agreement between the data and the model is thus compelling. However, with a considerable number of marginally constrained (and in some cases poorly understood) values at hand, it would not come as a surprise if the utilised model admitted significant variability in the distributions it could produce. Although a complete search of the parameter space is well beyond the scope of this study, it is noteworthy (as already alluded to above) that the model appears to be most sensitive to the orbital distribution of binary companions. While the median widest allowed binary orbit ($10^{3.75} \simeq 5\,600$~AU) in the simulation was motivated by the observational study of \citet{Kraus-etal-2011}, if we reduce this range to $10^{3\,}$AU, the enhancement near $\beta = 0$ in Figure~\ref{fig:Bat-MC} disappears entirely. On the contrary, if the range is extended to $10^{4.5\,}$AU, the enhancement near aligned orbits grows by almost a factor of $\sim 2$, although in both cases the shape of the PDF in the significantly misaligned region remains unchanged. The physical reason behind this is not exceedingly difficult to understand. As binary orbits get wider, the free precession (i.e., torquing) frequency of the disk decreases. This allows stars to adiabatically trail their disks for extended periods of time \citep{Batygin-2012}. In turn, this means that by the time the secular resonance encounter between the stellar spin-axis precession rate and the disk precession rate happens, the disk mass is systematically lower, leading to smaller excitation of misalignment.

Ideally, one would like to examine the sensitivity of the model to the assumed parameters in greater detail, although for such an activity to be meaningful, the underlying physics (particularly in the case of pre-main-sequence rotational evolution) must await some clarification. Therefore, we leave this exercise for a later study.

\subsection{Other mechanisms}

\subsubsection{Perturbations to the planetary orbit}

A few processes of formation of hot Jupiters have been proposed by several authors, who provide the expected distribution of the final inclination of the hot Jupiter with respect to its initial orbital plane. As already argued in Section~\ref{sec:proj}, this angle should be $\Psi$ in the end. We have collected these data from the papers, and derived the expected distribution of the projected spin-orbit angle $\beta$. We refer the reader to the original papers for a detailed explanation of how the distributions of $\Psi$ were derived by the authors. The data is originally in the form of a histogram $(\alpha_i,N_i)$, where $\alpha_i$ is the angle at the centre of the bin, and $N_i$ the number of planets with $\Psi$ in this bin. From this, we constructed $f=\sum_i f(\beta|\Psi=\alpha_i)N_i$, the distribution of $\beta$. We then bin $f$ using the same bins as in the original histogram. We checked that spreading the $N_i$ planets among angles different from $\alpha_i$ in the bin does not change the final histogram of $\beta$ significantly.

\begin{figure}
\includegraphics[width=\linewidth]{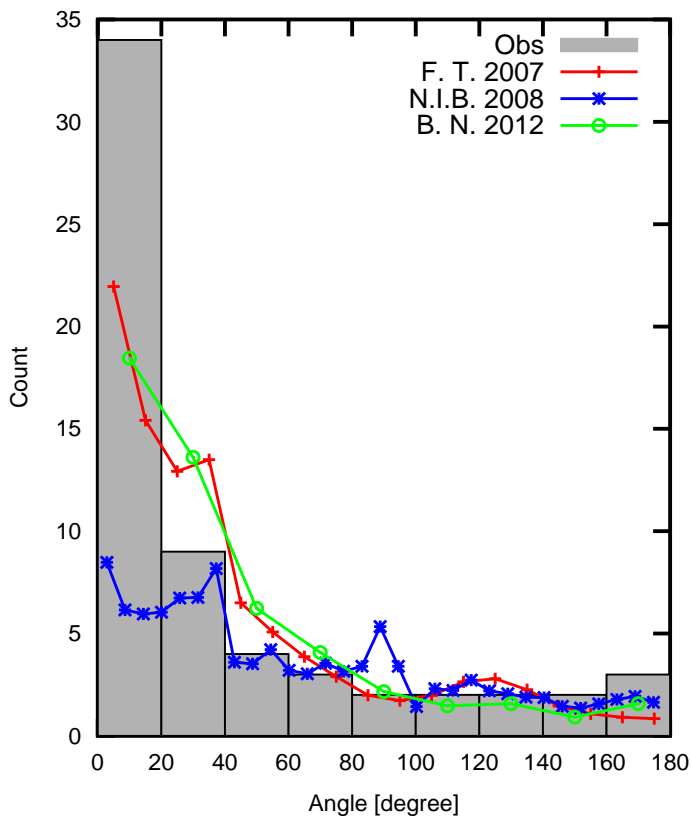}
\caption{Distribution of $\beta$ expected from several mechanisms of forming hot Jupiters. Background histogram\,: observations.}
\label{fig:All}
\end{figure}

The results are shown in Figure~\ref{fig:All}. The red line with \texttt{+} symbols labelled \texttt{F.T. 2007} represents the distribution of $\beta$ expected in the \citet{Fabrycky-Tremaine-2007} mechanism of Kozai cycles with tidal friction (their Fig.~10b providing $\Psi$ in bins of $10^\circ$). The blue line with stars labelled \texttt{N.I.B.2008} corresponds to the distribution found by \citet{Nagasawa-etal-2008} in their model of planet-planet scattering, tidal circularisation, and Kozai mechanism\,; the data is taken from their Figure~11c, which shows the histogram of $\Psi$ of formed close-in planets in all simulations, with bins of $0.1$~rad. The green line with circles labelled \texttt{B.N.2012} corresponds to the distribution found by \citet{Beauge-Nesvorny-2012} in their model of multi-planet scattering, in the case of three planets after 1 Gyrs of evolution (their fig.~16). All the distributions of $\beta$ have been normalised to have $18$ cases with $\beta>40^\circ$, for an easier comparison.

The agreement between the curves and the histogram beyond $40^\circ$
is in general satisfactory. Based on the small number of planets
observed in each bin, it seems impossible to exclude one or the other
mechanism at present. However, none of these processes can
individually account for the sharp peak at $\beta<20^\circ$. To this
end, \citeauthor{Beauge-Nesvorny-2012} point out that ``\emph{it is
  possible that the population of hot Jupiters with $\beta<20^\circ$
  have a different origin}''.

It should be noted, however, that \citet{Beauge-Nesvorny-2012} had
studied other scenarios, varying the inital number of planets, and the
final age of the systems. Four different cases are shown in their
paper, giving four different distributions. In particular, we have
checked that the case with four planets, after 3~Gyr of evolution,
once normalised to the $\beta>40^\circ$ cases, would give an excess of
aligned hot Jupiters compared to the observations. Thus, one can not
exclude that a combination of all the distributions matches the
observations. We found that their four published distributions give an
excess of planets in the $20^\circ<\beta<60^\circ$ range, but this
concerns only four cases. Actually, the observations do sample the
parameter space (in age and unknown initial conditions), so making a
combination would be appropriate.

\subsubsection{Perturbations to the stellar apparent spin}

Along a completely different line of thought, \citet{Rogers-etal-2012,Rogers-etal-2013} have proposed the misalignments to arise from the modulation of the outer layers of the host stars by interior gravity waves, rather than relics left behind by the dominant transport mechanism.

\citet{Cebron-etal-2011,Cebron-etal-2013} also suggest that the excitation of the elliptical instability in the star by the tides raised by the planet could give an apparent tilted rotation axis. Even the total spin of the star could change direction, if the coupling with the planet is efficient. This will also lead to misalignment between a star and its hot Jupiter.

In order to assess the viability of these intriguing ideas, an expected distribution of spin-orbit angles should be generated within the framework of these models, and compared against the observed distribution via a treatment such as that presented in this work.

\section{Discussion and conclusion}

In this paper, we provide a simple derivation of the probability
density function of the projected spin - orbit angle $\beta$, for
fixed real spin - orbit angle $\Psi$. This allows us to link models
(that produce distributions of $\Psi$) to observations (that measure
$\beta$).

Firstly, our geometric description shows that only positive values for
$\beta$ and $\Psi$ are sensible. These angles are between $0$ and
$180^\circ$, where $\Psi=0$ corresponds to prograde aligned orbits,
and $\Psi=180^\circ$ to retrograde aligned orbits. In particular, the
notation $\lambda=-\beta$ often found in the litterature is
irrelevant.

Second, we find that the observed distribution of $\beta$ presents a significant excess of quasi-aligned hot Jupiters ($\beta<20^\circ$) compared to the one expected from most models. This suggests that the scattering models and the Kozai cycle tidal friction models can not be solely responsible for the production of hot Jupiters. In previous studies \citep[see][]{Winn-etal-2010,Albrecht-etal-2012}, the excess of nearly-aligned hot Jupiters has been attributed to tidal re-alignment of the star. However, \citet{Rogers-Lin-2013} recently pointed out that tidal re-alignment preferentially leads to prograde aligned, retrograde aligned, or orthogonal spin-orbit angles, in some contradiction with the observed distribution \citep{Lai-2012}.

Third, in the simplest variant of the disk-torquing model, the over-representation of quasi-aligned planets is also not reproduced if the binaries responsible for the excitation of orbital obliquity have orbital semi-major axes smaller than $\sim 10^3$\,AU. However, within the framework of the picture envisioned by \citet{Batygin-Adams-2013arXiv}, alternative explanations are possible. These include adiabatic trailing of the host star, and early stripping or non-existance of the binary companion. In fact, the fraction of aligned disks increases dramatically with the maximum semi-major axes of the exciting binaries, $a_{\rm max}$. To this end, it is also worth noting that the wide binary fraction in star formation environments is a strong, growing function of primary stellar mass \citet{Kraus-etal-2011}. Thus, a thorough investigation of the mass-dependence of the disk-torquing mechanism appears worthwhile.

On the other hand, the expected distribution of $\beta$ for $\beta>40^\circ$ in the disk-torquing model hardly depends on $a_{\rm max}$, and is in very good agreement with the observations. Although other parameters, poorly constrained, can have an influence on the distribution of $\Psi$ and $\beta$ in the disk-torquing mechanism, we can conclude that the presently observed distribution of the spin - orbit angles is compatible with most hot Jupiters having been transported by smooth migration inside a proto-planetary disk, possibly torqued by a companion.

\subsection*{Acknowledgments}
K.B. acknowledges the generous support from the ITC Prize Postdoctoral
Fellowship at the Institute for Theory and Computation,
Harvard-Smithsonian Center for Astrophysics.  This research has made
use of the Exoplanet Orbit Database and the Exoplanet Data Explorer at
\texttt{exoplanets.org} \citep{exoplanets.org}. We thank the referee
C. Beaug\'e, as well as D. Nesvorny, for comments and suggestions that
led to improvement of this article. We further thank S. Tremaine for
having pointed out a mistake (now corrected) in our section 2.4.

\small

\bibliographystyle{aa}
\bibliography{/home/crida/MY_PAPERS/crida.bib}

\end{document}